\documentclass[%
 reprint,
superscriptaddress,
amsmath,amssymb,
aps,
pre,
floatfix,
]{revtex4-1}
\usepackage{enumitem}
\setlist[enumerate]{itemsep=0mm}
\usepackage{graphicx}% Include figure files
\usepackage{dcolumn}% Align table columns on decimal point
\usepackage{bm}% bold math
\usepackage[T1]{fontenc}
\usepackage[utf8]{inputenc}
\DeclareMathOperator*{\argmax}{arg\,max}

\usepackage{placeins}
\begin{document}

\title{Exploring the solution landscape enables more reliable network community detection}

\author{Joaqu\'\i{}n Calatayud}
\affiliation{Integrated Science Lab, Department of Physics, Ume\r{a} University, Sweden}
\author{Rub\'en Bernardo-Madrid}
\affiliation{
 Department of Conservation Biology, Estaci\'on Biol\'ogica de Do\~{n}ana (EBD-CSIC), Spain
 }%
\author{Magnus Neuman}
\author{Alexis Rojas}
\author{Martin Rosvall}
\affiliation{Integrated Science Lab, Department of Physics, Ume\r{a} University, Sweden}

\date{\today}

\begin{abstract}
To understand how a complex system is organized and functions, researchers often identify communities in the system's network of interactions. Because it is practically impossible to explore all solutions to guarantee the best one, many community-detection algorithms rely on multiple stochastic searches. But for a given combination of network and stochastic algorithm, how many searches are sufficient to find a solution that is good enough? The standard approach is to pick a reasonably large number of searches and select the network partition with the highest quality or derive a consensus solution based on all network partitions. However, if different partitions have similar qualities such that the solution landscape is degenerate, the single best partition may miss relevant information, and a consensus solution may blur complementary communities. Here we address this degeneracy problem with coarse-grained descriptions of the solution landscape. We cluster network partitions based on their similarity and suggest an approach to determine the minimum  number of searches required to describe the solution landscape adequately. To make good use of all partitions, we also propose different ways to explore the solution landscape, including a significance clustering procedure. We test these approaches on synthetic networks and a real-world network using two contrasting community-detection algorithms: The algorithm that can identify more general structures requires more searches and networks with clearer community structures require fewer searches. We also find that exploring the coarse-grained solution landscape can reveal complementary solutions and enable more reliable community detection.
\end{abstract}

\pacs{Valid PACS appear here}% PACS, the Physics and Astronomy
\maketitle

%\tableofcontents

\section{\label{sec:level1}Introduction}
Researchers in many disciplines across science use tools from network science to understand the structure, dynamics, and function of complex systems.
For example, identifying possibly nested groups of densely connected nodes, known as communities, with community-detection algorithms can highlight important network structures~\cite{girvan2002community,fortunato2010community,deritei2016principles,grilli2016modularity}. 
Most community-detection algorithms seek to find the set of communities, the network partition, that optimizes a quality score based on a specific definition of what constitutes a community.
Because finding the best network partition is an NP-hard problem, many algorithms rely on stochastic search strategies and require multiple runs to avoid local minima with bad solutions~\cite{rosvall2008maps,blondel2008fast,peixoto2014efficient}.
However, while they likely build communities from consistent small building blocks~\cite{riolo2019consistency}, all algorithms are more or less sensitive to degenerate solutions with similar quality scores for dissimilar partitions~\cite{good2010performance}. Moreover, small changes in an algorithm parameter \cite{weir2017post} or a network due to noise \cite{decelle2011inference} can drastically change the best solution, and a weak community structure can worsen this degeneracy problem. Therefore, reliable community detection must successfully deal with degenerate solutions.

\begin{figure}[hbtp]
\centering
\includegraphics[]{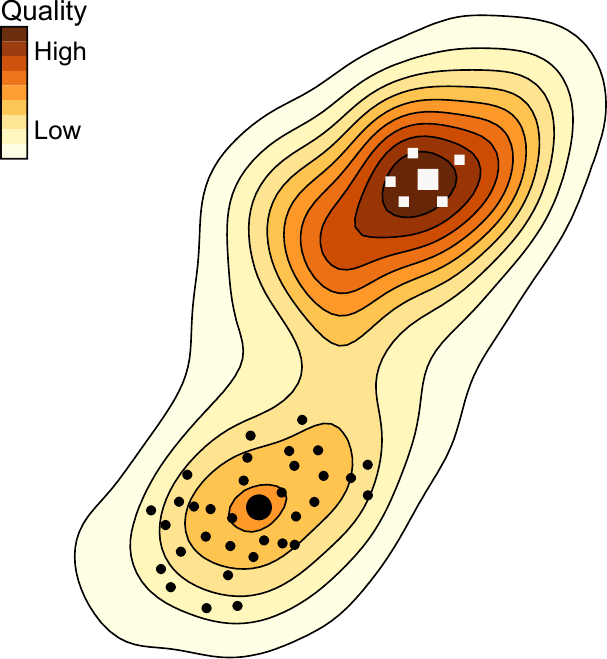}
\caption[]{A schematic solution landscape projected into a two-dimensional space with isolines for quality score. White squares and black circles represent two network partition clusters, with  partitions distributed based on their partition distances. Large symbols represent cluster centers. A consensus solution biased toward the numerous partitions marked with a black circle may have a lower quality score than any of the detected partitions.
}
\label{fig_1_ex} 
\end{figure}

To handle the degeneracy problem, consensus clustering seeks to combine information from multiple network partitions~\cite{strehl2002cluster,lancichinetti2012consensus,PhysRevE.99.042301}. The aim is to summarize the partitions in a single and possibly new partition with graph-based, combinatorial, or statistical techniques. Various approaches include finding the median partition or the one that shares the most information with other partitions~\cite{topchy2005clustering,strehl2002cluster}, consolidating groups of partitions with hypergraph methods~\cite{strehl2002cluster}, and re-clustering a co-occurrence network with the same community-detection algorithm \cite{lancichinetti2012consensus,PhysRevE.99.042301}. Although consensus clustering can alleviate some degeneracy problems and give higher quality solutions, using a single consensus partition can also waste important information or lead to misleading solutions if the partitions are incompatible. Moreover, disregarding the partition qualities can aggravate these problems when the number of low-quality partitions outweighs the number of high-quality partitions (Fig.~\ref{fig_1_ex}).

Studying the full solution landscape with all network partitions and corresponding quality measures results in no wasted information. However, such approaches are in practice limited to approximate visual explorations and the qualitative assessment of degenerate solutions~\citep{good2010performance,peixoto2017bayesian}. Moreover, for a given network and community-detection algorithm, it is unclear how many solutions are enough to describe the solution landscape adequately. As a result, we lack quantitative approaches that both highlight essential structures in the solution landscape and determine when it is safe to stop searching for better solutions. These challenges call for novel methods to comprehend and make use of the solution landscape to better understand the structure and dynamics of complex systems.

Here we present a partition clustering approach that explores the solution landscape of standard and hierarchical community-detection algorithms. To assess the completeness of the coarse-grained solution landscape, we cluster similar partitions together with a fast stream-clustering algorithm and estimate the probability that new partitions will fall within already defined partition clusters. For a coarse-grained solution landscape that meets a user-specified resolution, we propose different ways to explore the space of partitions, including visual explorations to reveal complementary solutions and a statistical test to identify significant communities. We validate our approach on synthetic networks as well as a real-world network of worldwide mammal occurrences.

\section{\label{sec_2}Describing the solution landscape}

\subsection{\label{sec_21}Network partition distance}
To describe the solution landscape, we first compute distances between partitions. While any of the many partition distance measures developed for different networks and research questions would work, most of them apply only to hard partitions that cannot capture hierarchical or overlapping community structures~\cite{danon2005comparing,hubert1985comparing,rand1971objective}. Because many real-world networks form these types of community structures~\cite{lancichinetti2009detecting,rosvall2011multilevel,perotti2015hierarchical,palla2005uncovering}, some distance measures have been generalized to capture either overlapping or hierarchical communities~\cite{lancichinetti2009detecting,perotti2015hierarchical,collins1988omega}, but rarely both~\cite{gates2019element}. To capture different types of community structures and make it easy to interpret the results, we want a flexible and simple distance measure.

Because a community of nodes is the building block of all types of community structures, we base the partition distance measure on pairwise community comparisons, regardless of whether they are in different hierarchical levels or whether nodes belong to more than one community. Specifically, we measure the weighted average of the minimum Jaccard distance over all communities in partition $P$ to a community in partition $P'$, with the weight given by the fraction of node assignments. That is, for each community $i$ in partition $P$ with set of nodes $C_i^P$, we measure the minimum Jaccard distance to any community $j$ in partition $P'$ with set of nodes $C_j^{P'}$, and calculate the weighted average based on the number of nodes in $C_i^P$, $|C_i^P|$, and the number of community assignments in partition $P$, $\sum_k |C_k^P|$, such that the distance $d_{PP'}$ from partition $P$ to partition $P'$ is
\begin{equation}\label{np-distance}
	d_{PP'}=\sum_{i}\underset{j}{\textrm{min}}\left( 1-\frac{|C_i^P\cap{C_j^{P'}}|}{|C_i^P\cup{C_j^{P'}}|}\right)\frac{|C_i^P|}{\sum_k |C_k^P|}.
\end{equation}
Because $d_{PP'}$ need not be equal to $d_{P'\!P}$, for a symmetric partition distance measure, we calculate the average~\cite{goldberg2010measuring},
\begin{equation}
	\bar{d}_{PP'}=\frac{1}{2}d_{PP'}+\frac{1}{2}d_{P'\!P}.
\end{equation}
This partition distance works with hard, overlapping, and hierarchical communities. It is zero for identical partitions, and approaches 1 as they become completely dissimilar. Between these extremes, the partition distance gives the weighted average fraction of nodes that best-matching communities do not have in common.

\subsection{\label{sec_22}Network partition clustering algorithm}

Using the proposed network partition distance, we describe the solution landscape with clusters of similar network partitions. While many clustering algorithms can output such clusters, those algorithms generally involve NP-hard optimization problems in themselves. However, to identify dissimilar partitions with high quality, we do not need a solution landscape that optimizes some quality function. Instead, a fast and transparent deterministic approach that decides the number of clusters provides multiple advantages: First, a fast algorithm can run together with a stochastic community-detection algorithm and decide when it is safe to stop to achieve a good result. Second, a deterministic algorithm that does not require a prespecified number of clusters evades the ambiguities that come with multiple solutions. Third, a transparent algorithm that produces interpretable clusters and a comprehensible solution landscape simplifies further analysis. Therefore, instead of relying on established clustering algorithms developed for other purposes, given a partition distance threshold $d_{\text{max}}$, we cluster the partitions in three steps:
\begin{enumerate}
    \item Order all $p$ network partitions from highest to lowest quality.
    \item Let the highest quality network partition form cluster center 1.
    \item Repeat until all network partitions have been clustered. Among the not-yet-clustered partitions, pick the one with the highest quality and assign it to the first of the $k$ cluster centers that it is closer to than $d_{\text{max}}$. If no such cluster center exists, let it form cluster center $k+1$.
\end{enumerate}

For example, in the schematic solution landscape in Fig.~\ref{fig_1_ex}, the network partition clustering algorithm first lets the partition marked with a big square form the center of cluster 1. For an intermediate partition distance threshold, it then assigns the other partitions marked with squares to the same cluster before it lets the partition marked with a big circle form the center of cluster 2 and assigns the other partitions marked with circles to that cluster. 

The partition distance threshold specifies the resolution of the coarse-grained solution landscape. Lowering the threshold gives more clusters with more similar network partitions and increasing the threshold gives fewer clusters with less similar network partitions.

We have implemented the partition clustering code in C++, which has worst-case time-complexity $\mathcal{O}(pk)$, and made it available for anyone to use at \url{https://github.com/mapequation/partition-validation}

\subsection{\label{sec_24}Solution landscape completeness}

We say that a solution landscape is complete when new network partitions at most marginally affect its coarse-grained description. Accordingly, when a solution landscape is complete, it is safe to stop searching for better network partitions. Intuitively, we need fewer partitions to describe the solution landscape of a network with a clear community structure than that of a network with a diffuse community structure because the former will have more similar partitions. Moreover, the required number of partitions will also depend on the variability of the search algorithm. In any case, using more partitions to describe the solution landscape with clusters increases the probability that a new partition will fit into existing clusters. We use this probability as a validation score $\sigma$ to assess the solution landscape completeness and to determine when to stop searching, 
\begin{equation}
    \sigma=\frac{p_c}{p_v},
\end{equation}
where $p_c$ is the number of validation partitions that fits within a cluster and $p_v$ is the total number of validation partitions. For example, we can stop the search algorithm when $\sigma$ is higher than, say, 0.9. To estimate $\sigma$, we use repeated random sub-sampling validation and hold out $p_v=100$ partitions for validation, or $p_v=p/2$ when the number of partitions is fewer than 200. In this way, we avoid random effects caused by the search order of the stochastic community-detection algorithm.

\subsection{\label{sec_23}Solution landscape exploration}

A complete coarse-grained solution landscape with clusters centered around locally high-quality partitions simplifies further analysis and makes the results more reliable. First, it indicates when it is safe to stop searching for a better solution because the validation score and partition distance threshold put a limit on the value of continuing. For example, when a solution landscape is complete at a high validation score for a small partition distance threshold, summary statistics based on all partitions will be reproducible and reliable. Second, it directly gives an idea about the spread of network partitions through the number of clusters for a given partition distance threshold. For more detailed analysis, alluvial diagrams can highlight qualitative pairwise differences between partitions~\cite{rosvall2010mapping} and various embedding techniques can depict the overall solution landscape~\cite{maaten2008visualizing}. Third, it can speed up further analysis with controlled information loss as comparing all pairs of cluster centers rather than all pairs of partitions reduces the computational complexity from $\mathcal{O}(p^2)$ to $\mathcal{O}(k^2)$.

Useful further analysis include finding communities or node assignments that are stable across many partitions. For example, in networks with partially clear community structure, distinguishing stable from unstable communities enables more reliable analysis. While approaches exist for assessing the significance of communities given a set of partitions~\cite{karrer2008robustness,rosvall2010mapping}, these approaches only work for hard non-hierarchical partitions. Therefore, we propose an approach that also assesses the significance for hierarchical or overlapping communities. A straightforward approach to assessing the significance of a community would be to calculate the fraction of partitions in which the community appears. However, this significance test is overly demanding as communities with only slight variations in node composition would be considered non-significant. Consequently, we relax the demand for exact matching and reuse the minimum Jaccard distance of the network partition distance in Eq.~(\ref{np-distance}) with a threshold. We measure the significance $\alpha_i^R$ of community $i$ in the highest-quality or other reference partition $R$ as the fraction of partitions that have a community with a smaller distance to $i$ than a threshold $\tau$,
\begin{equation}
	\alpha_i^R=\frac{1}{p-1}\sum_{P \neq R}\Theta\left[\tau - \min_j\left( 1-\frac{|C_i^R\cap{C_j^P}|}{|C_i^R\cup{C_j^P}|}\right) \right],
\end{equation}
where the sum runs over all $p-1$ partitions $P$ that are not the reference partition $R$ and $\Theta$ is the Heaviside step function.

Stable communities can contain both stable and unstable node assignments, and we need a means to distinguish between them. Therefore, to measure the community-assignment significance $\eta_v^R$ of node $v$ in reference partition $R$, we calculate the fraction of partitions in which $v$ appears in the community that is most similar to $v$'s community in the reference partition. Using the Kronecker delta function $\delta$, the community-assignment significance can be written
\begin{equation}
	\eta_v^R=\frac{1}{p-1}\sum_{P \neq R}\delta\left(c_v^P,c_v^{RP}\right),
\end{equation}
where $c_v^P$ is the community index of node $v$ in partition $P$, and ${c_v^{RP} = \argmax_{j} |C_{c_v^R}^R\cap{C_j^P}|/|C_{c_v^R}^R\cup{C_j^P}}|$ is the community index of the community in partition $P$ that is most similar to the community of $v$ in partition $R$. In practice, we calculate $\eta_v^R$ in four steps:
\begin{enumerate}
    \item Identify the index $c_v^R$ of $v$'s community in the reference partition.
    \item Identify the index $c_v^{RP}$ of the community in partition $P$ that is most similar to community $c_v^R$ in the reference partition.
    \item Increment $\eta_v^R$ by $1/(p-1)$ if the index $c_v^P$ of $v$'s community in partition $P$ is the same as the most similar partition $c_v^{RP}$.
    \item Repeat 2.\ and 3.\ for all $p-1$ partitions $P$ that are not the reference partition.
\end{enumerate}

\section{\label{sec_3}Results and Discussion}
\subsection{\label{sec_31}Solution landscape of synthetic networks}
We tested our approach on Lancichinetti-Fortunato-Radicchi (LFR) benchmark networks with different intercommunity link probabilities $\mu$~\cite{lancichinetti2008benchmark}. We generated networks with 500 nodes, average degree 10, maximum degree 20, community sizes distributed between 20 and 100 nodes, and four different intercommunity link probabilities $\mu=0.1$, 0.15, 0.2, and 0.25,  for less and less pronounced communities. To account for the internal variability of the LFR benchmark networks, we generated 25 synthetic networks for each $\mu$.

We analyzed these networks with two popular and contrasting stochastic algorithms for community detection: Infomap~\cite{rosvall2008maps,infomap} and Bayesian inference of the degree-corrected stochastic blockmodel (BSBM)~\citep{peixoto2017nonparametric} as implemented in the graph-tool library~\citep{peixoto2014efficient,peixoto2014graph}. While both algorithms optimize information-theoretic objective functions, Infomap seeks to compress dynamics on a network with assortative communities of densely connected nodes whereas the BSBM seeks to compress the network itself with blocks of any mixing pattern.  Moreover, the BSBM can handle partition uncertainty based on sampling from the posterior distribution~\citep{peixoto2017bayesian}. To test the solution landscape completeness, we ran each algorithm 50, 100, 200, 300, 400, and 500 times on a given network. After each step, we ran the partition clustering algorithm and validated 100 times on 100 sub-sampled hold-out partitions when $p\geq200$ and on $p/2$ partitions otherwise. 

\begin{figure}[tb] 
\includegraphics[]{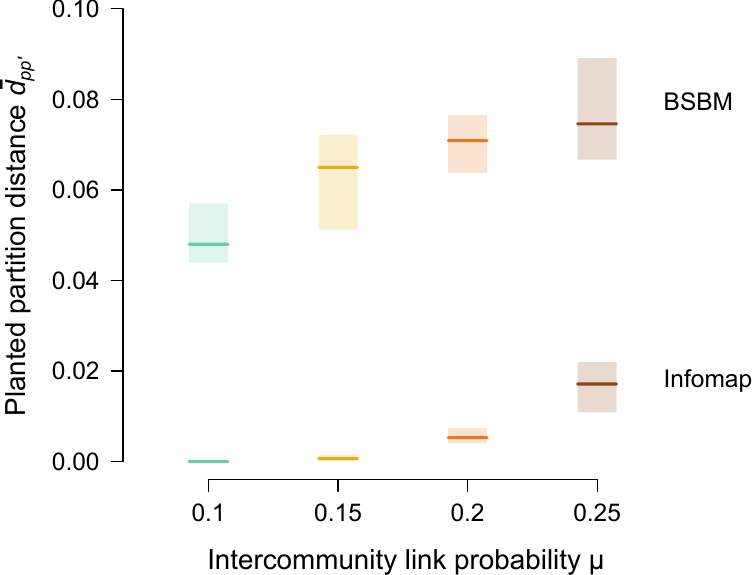} 
\caption[]{Distance from Infomap and BSBM partitions to planted partitions obtained with the LFR benchmark. Dark medians and light bars between the 25th and 75th percentiles over 500 partitions for each intercommunity link probability $\mu$.}
\label{fig_2}
\end{figure}

\begin{figure}[tb] 
\includegraphics[width=\columnwidth]{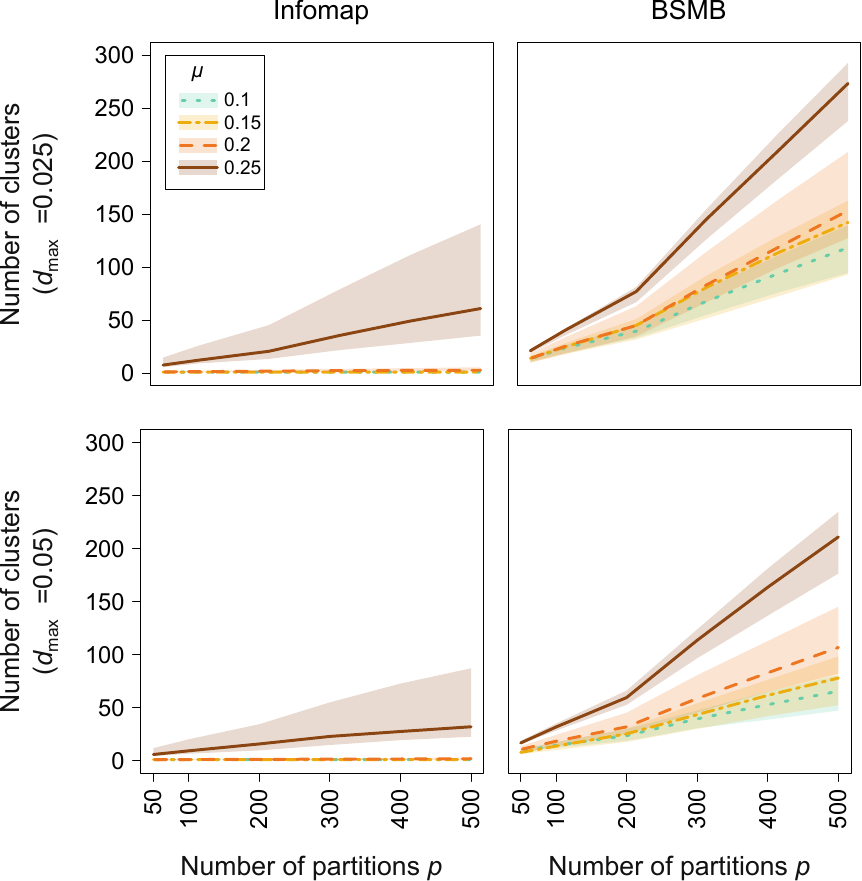} 
\caption[]{The number of clusters describing Infomap and BSBM solution landscapes for synthetic networks generated with four intercommunity link probabilities $\mu$. Solid medians and shaded regions between the 25th and 75th percentiles for 50--500 partitions with partition distance thresholds $d_{\text{max}} = 0.025$ and $d_{\text{max}} = 0.05$. The more variable BSBM partitions require more clusters.}
\label{fig_3}
\end{figure}

\begin{figure}[tb] 
\includegraphics[width=\columnwidth]{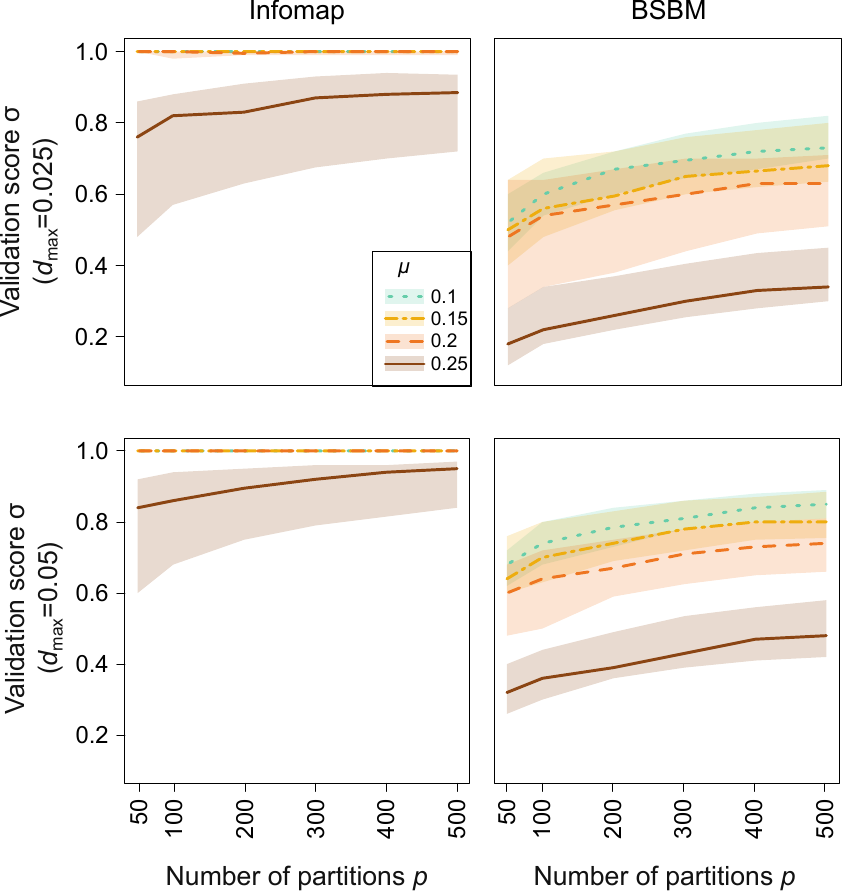}
\caption[]{Infomap and BSBM solution landscape completeness for synthetic networks generated with four intercommunity link probabilities $\mu$. The validation scores $\sigma$ with solid medians and shaded regions between the 25th and 75th percentiles for different numbers of partitions with partition distance thresholds $d_{\text{max}} = 0.025$ and $d_{\text{max}} = 0.05$. Infomap requires fewer partitions than the BSBM for complete solution landscapes. Both methods require more partitions for less pronounced communities.}
\label{fig_4}
\end{figure}

With the more general model not limited to assortative communities and the Bayesian framework, the BSBM has a flatter solution landscape than Infomap. As a result, the BSBM generated more variable partitions that differed more from the planted partitions (Fig.~\ref{fig_2}) and required more clusters for any tested intercommunity link probability and distance threshold (Fig.~\ref{fig_3}). While both methods required more partitions for networks with higher intercommunity link probabilities $\mu$ -- such that a less pronounced community structure required a larger number of searches to obtain a complete solution landscape -- Infomap generated partitions with validation scores close to 1 for $\mu \le 0.2$ already at 50 partitions (Fig.~\ref{fig_4}).

The required number of searches also depends on the choice of partition distance threshold $d_{\text{max}}$. To exemplify this, we used two threshold values for validation, $d_{\text{max}} = 0.025$ and $d_{\text{max}} = 0.05$. With the higher threshold, more hold-out partitions fit in clusters such that the validation score increases (Fig.~\ref{fig_4}). Therefore, the choice of partition distance threshold should reflect a compromise between accuracy and efficiency and depend on the particular problem at hand, which we exemplify in the next section.

\subsection{\label{sec_32}Solution landscape of a mammal occurrence network}
We further explored the solution landscape in a real-world case using a terrestrial mammal occurrence network. This bipartite network consists of 4999 mammal species and 10,775 equal-area grid cells with 110.5 km sides that cover the surface of the Earth~\cite{Bernardo-Madrid287300}. A link connects a species and a grid cell if the species occurs in the grid cell. The resulting communities form global-scale areas that share similar species called bioregions.

We analyzed the community structure with the hierarchical versions of Infomap~\cite{rosvall2011multilevel} and the BSBM~\cite{peixoto2014hierarchical} by generating 1500 partitions with each algorithm. We chose $d_{\text{max}}=0.2$, which roughly corresponds to partition differences that cover up to 20\% of the Earth's surface. Higher partition distances indicate major changes in the bioregional configuration, which require separate examination. Nevertheless, to illustrate the effect of different thresholds, we used three smaller values, $d_{\text{max}}=$ 0.025, 0.05, and 0.1. To validate the solution landscape under different numbers of runs, we used 200--1500 partitions with 100 hold-out partitions sub-sampled 100 times.

\begin{figure}[tbp] 
\includegraphics[]{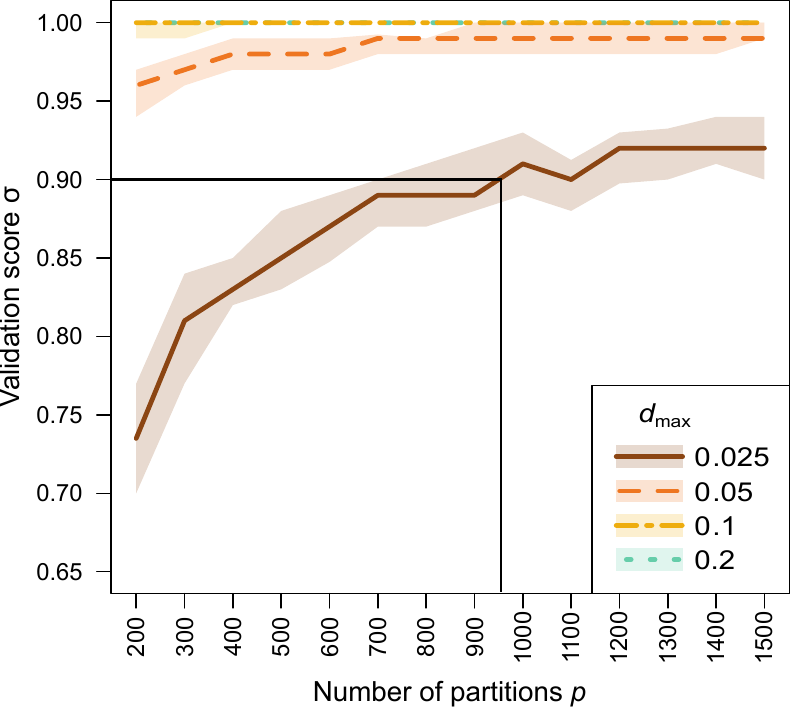}
\caption[]{Validation score $\sigma$ for landscape completeness of the terrestrial mammal occurrence network under four partition distance threshold values $d_{\text{max}}$ (0.2, 0.1, 0.05, 0.025) using Infomap. One thousand partitions are enough for a median validation score above 0.9.}
\label{fig_5}
\end{figure}

The results on the real networks resemble those on the synthetic networks. Compared with Infomap, the BSBM again generated more variable partitions and a more complex solution landscape. Because the distance was higher than $d_\mathrm{max}=0.2$ between each pair of the BSBM partitions, each partition formed its own cluster such that the validation score $\sigma$ was 0. For distance threshold $d_\mathrm{max}=0.2$, we would need vastly more than 1500 partitions to achieve $\sigma=0.9$. While we obtained $\sigma\simeq0.9$ with 1500 partitions by increasing $d_\mathrm{max}$ to $0.55$, this distance threshold allows overly dissimilar partitions: two partitions in the same cluster can have best-matching communities that on average share less than half of their nodes. Accordingly, many different block structures can generate this network with similar probabilities. Focusing on assortative communities simplifies the problem: With Infomap we achieved complete solution landscapes with $\sigma > 0.9$ for all tested threshold values $d_{\text{max}}$ (Fig.~\ref{fig_5}). For example, for the lowest tested $d_{\text{max}}$ = 0.025, $\sigma$ was higher than 0.9 when we used more than 900 partitions, which formed 188 clusters (Fig.~\ref{fig_5}). In contrast, for the highest tested $d_{\text{max}} = 0.2$, $\sigma$ was higher than 0.9 already at 200 partitions (and likely before), and the 1500 partitions formed two clusters with 970 and 530 partitions, respectively. The cluster centers have similar qualities, 10.689 and 10.695, which Infomap measures as code lengths in bits. Indeed, the clusters have partitions with overlapping code lengths (from 10.695 and 10.697 at the 25th percentile to 10.700 for both clusters at the 75th percentile), which call for further analysis of the degenerate solution landscape.

\begin{figure*}[bt] 
\includegraphics[]{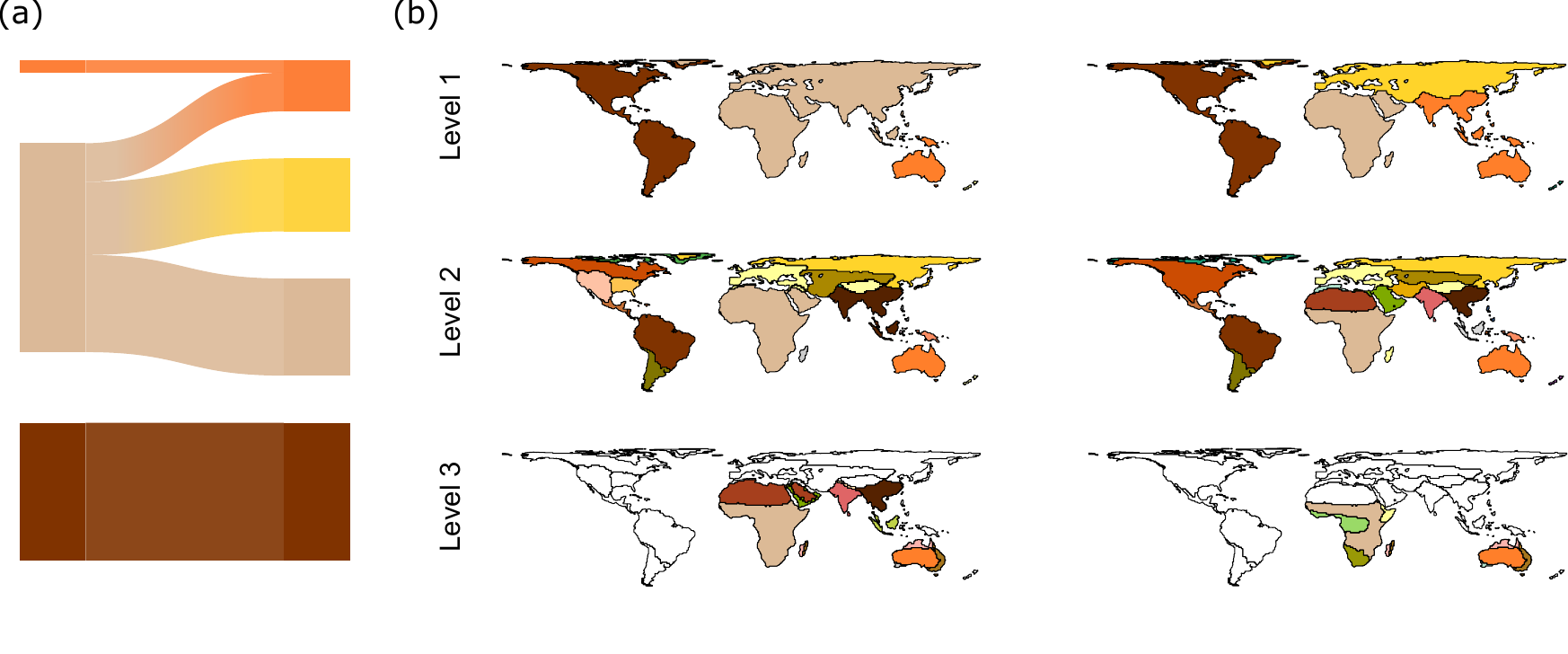}
\caption[]{World bioregions from communities in the two best partition cluster centers.(a) Alluvial diagram showing the differences between the two partitions at the highest hierarchical level. (b) Geographic projection of nodes representing grid cells. In all cases, we obtained three hierarchical levels. The differences show the rich information contained in separate partitions, even when they have similar quality.}
\label{fig_6}
\end{figure*}

To explore the qualitative differences between the clusters, alluvial diagrams can give a visual overview of major changes between the cluster centers (Fig.~\ref{fig_6}(a)). In our particular case, however, we can visualize the geographic projection of the spatially explicit grid cells (Fig.~\ref{fig_6}(b)). At the highest hierarchical level (level 1 in Fig.~\ref{fig_6}(b)), the major difference is that the second cluster center  splits Africa and a southeastern portion of Asia from a large region that encompasses Eurasia and Africa in the first cluster center. At lower hierarchical levels, the first cluster center further subdivides the North American region whereas the second cluster center further subdivides regions in Africa and central Asia. These results show the rich information contained in different partitions, which can reveal meaningful patterns. For instance, the subdivision of Sub-Saharan Africa closely coincides with the K\"{o}ppen climate classification~\cite{kottek2006world}.

\begin{figure*}[tb] 
\includegraphics[]{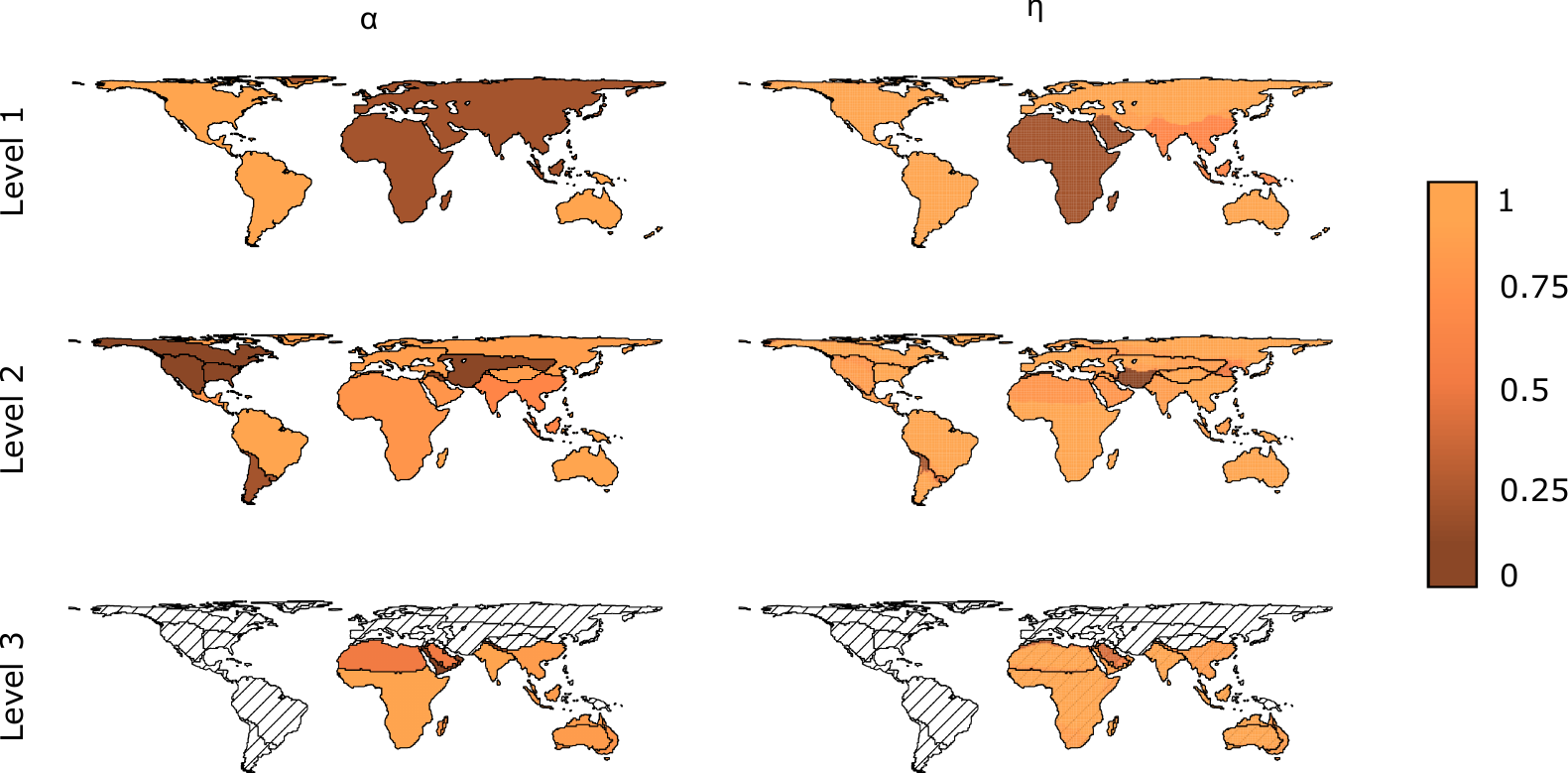}
\caption[]{The fraction of partitions having a community no more distant than $\tau=0.2$ to the reference community (left). The fraction of partitions where a node belongs to the most similar community (right). We see for example that the weakly supported African Euro-Asiatic region in the first level appears to hold a significant core of nodes coinciding with the north of Eurasia, while less significant nodes tend to be placed along bioregional borders. The striped areas correspond to regions that were not further subdivided in the third hierarchical level.}
\label{fig_7}
\end{figure*}

Finally, we applied the significance clustering procedure both at the community and node level with the overall highest quality partition as a reference. We used community distance threshold $\tau = 0.2$ to calculate the community significance $\alpha^R_i$. The community significance is largely in agreement with the previous qualitative visual assessment. The region including Africa and Eurasia is weakly supported, which is also true for the North American and Central Asian regions (Fig.~\ref{fig_7}). Also, the node significance $\eta^A_i$ agrees with these results, but provides further information. For instance, the weakly supported African Euro-Asiatic region in the first level appears to hold a significant core of nodes coinciding with northern Eurasia. Moreover, nodes with low significance tend to be placed along regional borders such as the Sahel border and the border separating southern and northern South America. Beyond methodological stochasticity, this result shows that some nodes are inherently more difficult to assign to particular communities.

\section{\label{sec_4}Conclusions}
We have introduced a fast network partition clustering algorithm to describe the often degenerate solution landscape of stochastic community-detection algorithms in coarse-grained form. Our approach establishes a criterion for when it is safe to stop searching for better solutions and start exploring the solution landscape. We also present new statistical tests of communities and node assignments, which uncover underlying causes of the solution landscape degeneracy. The validation on synthetic networks and a real-world network highlights how focusing on a single network partition can waste useful information. In contrast, using the entire solution landscape enables more reliable community detection and a better understanding of the organization of complex systems. Beyond community detection, our approach works with any stochastic search with outputs that have measurable distances.

\begin{acknowledgements}
We thank Eloy Revilla for stimulating discussions, Tiago Peixoto for explaining BSBM behavior, and Anton Eriksson for constructing the alluvial diagram in Fig.~\ref{fig_6}. We are also grateful to Juan Ignacio Perotti for detecting a typo in the equations. J.C. was supported by the the Carl Trygger Foundation, R.B.-M. by the Spanish Ministry of Economy, Industry and Competitiveness, grant BES-2013-065753, M.R.\ by the Swedish Research Council, grant 2016-00796, and M.N. and A.R. were supported by the Olle Engkvist Byggm\r{a}stare Foundation.
\end{acknowledgements}

%\bibliography{biblio.bib}
\FloatBarrier
\providecommand{\noopsort}[1]{}\providecommand{\singleletter}[1]{#1}%

\end{document}